\documentclass[prb,twocolumn,floatfix,showpacs,amsmath,amssymb,aps,superscriptaddress]{revtex4-1}
\usepackage{graphicx}
\usepackage{amsmath}
\usepackage{subfigure}
\usepackage{times}
\usepackage{xcolor}
\usepackage{comment}

\usepackage{color}
\usepackage[colorlinks,citecolor=blue]{hyperref}

\newcommand{\beq}{\begin{equation}}
\newcommand{\eeq}{\end{equation}}
\newcommand{\bea}{\begin{eqnarray}}
\newcommand{\eea}{\end{eqnarray}}
\newcommand{\bpm}{\begin{pmatrix}}
\newcommand{\epm}{\end{pmatrix}}
\newcommand{\bef}{\begin{figure}}
\newcommand{\eef}{\end{figure}}

\begin{document}
\title{Topological crystalline semimetals in non-symmorphic lattices}
\author{Yige Chen}
\affiliation{Department of Physics, University of Toronto, Ontario M5S 1A7 Canada}

\author{Heung-Sik Kim}
\affiliation{Department of Physics, University of Toronto, Ontario M5S 1A7 Canada}

\author{Hae-Young Kee}
\affiliation{Department of Physics, University of Toronto, Ontario M5S 1A7 Canada}
\affiliation{Canadian Institute for Advanced Research, CIFAR Program in Quantum Materials, Toronto, ON M5G 1Z8, Canada}
\email{hykee@physics.utoronto.ca}

\date{\today}
\begin{abstract}
Numerous efforts have been devoted to reveal exotic semimetallic phases with topologically non-trivial bulk and/or surface states in materials with strong spin-orbit coupling.
In particular, semimetals with nodal line Fermi surface (FS) exhibit novel properties, and searching for candidate materials becomes an interesting research direction.
Here we provide a generic condition for a four-fold degenerate nodal line FS in non-symmorphic crystals with inversion and time-reversal symmetry (TRS). 
When there are two glide planes or screw axes perpendicular to each other, a pair of Bloch bands related by non-symmorphic symmetry
become degenerate on a Brillouin Zone (BZ) boundary.
There are two pairs of such bands, and they disperse in a way that the partners of two pairs are exchanged on other BZ boundaries. 
This enforces a nodal line FS on a BZ boundary plane protected by non-symmorphic symmetries. 
When TRS is broken, four-fold degenerate Dirac points or Weyl ring FS could occur depending on a direction of the magnetic field. 
On a certain surface  double helical surface states exist, which become double Ferm arcs as TRS is broken. 
\end{abstract}
\maketitle

\section{Introduction}
Recently intense interest has been drawn to novel topological semimetallic phases, in which the systems support non-trivial band crossing points in crystal momentum space. Such studies have been motivated by the discovery of topological insulators with bulk energy gap and conducting surface modes protected by TRS ~\cite{KanePRL05,BernevigPRL06,BernevigSCI06,FuPRB07,FuPRL07,MoorePRB07,KonigSCI07,HsiehNature08,HsiehSCI09,HasanRMP10,QiRMP11}. 
A list of topological semimetals, which is an extension of topological insulators to metallic phases, has been growing in theoretical communities, and some members in the list have been experimentally confirmed.\cite{LiuSCI14, Gibson14,XuSCI15,LvPRX15}
One group of topological semimetals is characterized by FS points. This includes Weyl semimetal with chiral fermion ~\cite{Volovik07,Murakami07,BurkovPRL11,WanPRB11,YangPRB11}, and three-dimensional (3D) Dirac semimetals with surface Fermi arc states~\cite{YoungPRL12,WangPRB12,WangPRB13}.
Another class of topological semimetals is characterized by a closed loop of FS called nodal line FS
~\cite{Carter12,BurkovPRB11,ChiuPRB14,Phillips14,Weng14,Kim15,Chen15,Chiu15,Bian15,Zeng15}. These semimetals named as topological nodal line semimetals have recently been proposed in various materials, including a three-dimensional graphene network~\cite{Weng14}, Ca$_3$P$_2$~\cite{XieAPL15}, Cu$_3$PdN~\cite{YuPRL15}, and orthorhombic perovskite iridates~\cite{Carter12,Chen15}. However, in graphene material, Ca$_3$P$_2$, and Cu$_3$PdN, spin-orbit coupling gaps out the nodal FS, and the system becomes a trivial insulator~\cite{Zeng15}. On the other hand, in perovskite iridates, spin-orbit coupling assists the system to develop nodal line FS~\cite{Carter12}.

In this work, we provide a generic condition for a four-fold degenerate nodal line FS for three-dimensional spin (or pseudospin)-1/2 systems, where the non-symmorphic crystal symmetry plays a critical role.
In the presence of space-time inversion symmetry, all Bloch states are doubly degenerate -- Kramers theorem. 
We show that when there are two glide planes or screw axes perpendicular to each other, two Bloch bands related by these operations form a degenerate pair on a BZ boundary.
Their Kramers partners follow the same pattern, and thus four Bloch states are degenerate on a BZ boundary.
Due to the two non-symmorphic symmetry operations, two pairs of such bands exist, and  
their partners within each pair should be exchanged when the Bloch bands disperse from one BZ boundary to another. 
This enforces a four-fold degenerate nodal line FS protected by non-symmorphic symmetries on a BZ boundary plane. 
Using a tight binding model derived for perovskite iridates $A$IrO$_3$ ($A$ being alkali earth elements)\cite{Carter12}, we further show how Dirac point or line of Weyl FS appears when TRS is broken. 
Surface states of this topological {\it crystalline} semimetal with and without TRS are also presented.

\section{A sufficient condition for nodal-line semimetal with TRS}
First we provide a generic condition for a four-fold degenerate nodal line FS as shown in Fig. 1 (a) : a combination of two perpendicular non-symmorphic symmetry operations together with space-time inversion symmetry 
guarantees a four-fold degenerate nodal line FS for three-dimensional spin (pseudospin)-1/2 system. 
Let us consider the $\hat{a}$- and $\hat{b}$-axis two-fold screw operators that are perpendicular to each other. 
Explicit form of those operations are given as follows.
\bea
\hat{S}_a: \left(x,y,z,t \right) &\rightarrow& \left(\frac{1}{2} +x, \frac{1}{2} -y, -z, t \right) \times i \hat{\sigma}_x, \label{eq:1:a}\\
\hat{S}_b: \left(x,y,z,t \right) &\rightarrow& \left(\frac{1}{2} - x, \frac{1}{2} +y, \frac{1}{2} -z,t \right) \times i \hat{\sigma}_y,\label{eq:1:b}
\label{eq:1:c}
\eea
where 
the Bravais lattice vectors $\vec{R}=x \vec{a}+y \vec{b} +z\vec{c}$.
We set the length of each lattice vector to unity, {\it i.e} $|\vec{a}|=|\vec{b}|=|\vec{c}|=1$. 
The Pauli matrices $\hat{\vec{\sigma}}=\left(\hat{\sigma}_x,\hat{\sigma}_y,\hat{\sigma}_z\right)$ represent how spin transforms under the above symmetry operations. 
Note that another screw-axis operator ${\hat S}_c$ is defined via ${\hat S}_c ={\hat S}_a *{\hat S}_b$.

In addition to these crystalline symmetries, the system also preserves time-reversal $\hat{T}$ and inversion $\hat{P}$ symmetries. The composite symmetry operator 
defined as the product of time-reversal and inversion operators ($\Theta\equiv\hat{T}*\hat{P}$) reverses the space-time and spin coordinates simultaneously,
$\Theta: \left(x,y,z,t \right) \rightarrow \left(-x,-y,-z,-t \right) \times i \hat{\sigma}_y\; $.
Since $\Theta^2=-1$, it enforces twofold degeneracy everywhere in the momentum space.

Note that the glide planes are found by taking the product of the above screw and inversion operators, i.e.,  $\hat{b}$-glide operator ${\hat G}_b = {\hat S}_a * {\hat P}$, 
n-glide ${\hat G}_n = {\hat S}_b * {\hat P}$, and mirror reflection at  $z=1/4$, ${\hat M}_c = {\hat S}_c * {\hat P}$. This corresponds to ${\rm P}_{bnm}$ lattice,
but our proof below is general and applicable to other lattices with two orthogonal non-symmporhic symmetries such as ${\rm P}_{bca}$. 

Now let us focus on the $k_b=\pi$ plane which is invariant under $\hat{G}_n$ operation. The Bloch states $|\phi_i \rangle$ on the plane carry $\hat{n}$-glide eigenvalues $n_{\pm}=\pm i e^{i \frac{k_a+k_c}{2}}$. Its Kramers partner, $\Theta |\phi_i \rangle$ is also an eigenstate of $\hat{G}_n$ with the same $\hat{n}$-glide eigenvalue on this BZ boundary plane\cite{FangPRB15}. 
There are eight Bloch states say  $|\phi_i\rangle$ and $\Theta |\phi_i \rangle$ where $i=1 \cdots 4$ for a single orbital problem.
In general, these Bloch states at a generic momentum point on $k_b=\pi$ plane carry $n_+$ and $n_-$ eigenvalues as shown in Fig. 1(b).
At a time-reversal invariant momentum (TRIM) point ${\rm U}=(k_a=0,k_b=\pi,k_c=\pi)$, these Bloch states can be classified by eigenvalues of the $\hat{n}$-glide as well as
a screw axis operator $\hat{S}_a$ along $\hat{a}$-axis.
Since $\left(\hat{S}_a\right)^2=-e^{i k_a}$, the eigenvalues $a_{\pm}=\pm i$ at U-point. 
Due to the presence of the $\hat{b}$-glide symmetry, two Bloch states $|\phi_{1}\rangle$ and $|\phi_{3} \rangle$, and $|\phi_2\rangle$ and $|\phi_4\rangle$ are in fact related under $\hat{G}_b$ {\it i.e.} $|\phi_{3}\rangle \propto \hat{G}_b|\phi_{1}\rangle$ and $|\phi_{4}\rangle \propto \hat{G}_b|\phi_{2}\rangle$ with the same $\hat{n}$-glide but opposite screw eigenvalues as denoted in Fig. 1(b) at the U-point. 
The four-fold degeneracy including Kramers partner at U-point is protected by the $\hat{n}$-glide and screw axis $\hat{S}_a$.  
On the other hand, along the R-S and X-S BZ boundary lines where the Bloch states within each pair are related though $\hat{G}_b$ and $\hat{S}_a$ respectively,
 $|\phi_{1}\rangle$ and $|\phi_{2}\rangle$, and $|\phi_{3}\rangle$ and $|\phi_{4}\rangle$ are degenerate with different $\hat{n}$-glide eigenvalues as shown in Fig. 2(c). 
 Thus the two pairs of Bloch states at U-point must experience a partner switching between $|\phi_2\rangle$ and $|\phi_3\rangle$ when the bands evolve towards the BZ boundary line X-S or R-S on $k_b=\pi$ plane.
This also occurs for Kramers partner states.
We refer the Supplemental Material for proofs of Bloch states degeneracies at different BZ boundaries originated from the glide/screw symmetries. This enforces a four-fold degenerate nodal line on $k_b=\pi$ plane with U-point as its center. The nodal line crossings hence are assured by the non-symmorphic space group at a half filling, which indicates the system should be a filling enforced semimetal\cite{SidNP13,Watanabe15}. 

\begin{figure}[t]
\centering
\includegraphics[width=9cm]{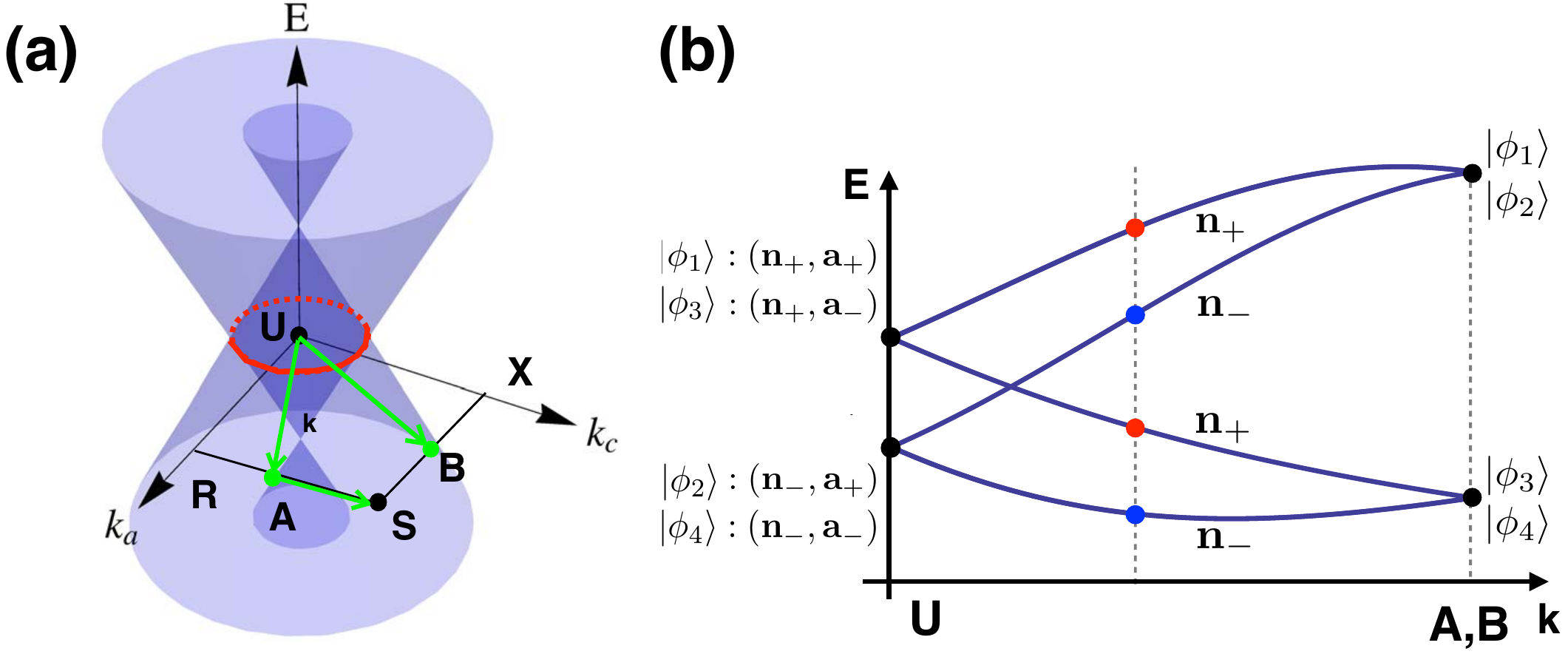}
\caption{
(color online) The four-fold degenerate nodal line on $k_b=\pi$ plane is shown as the red loop in (a). 
(b) shows the schematic band dispersion along the U-A and U-B lines, where A and B are arbitrary points on the R-S and X-S 
lines, respectively as depicted in (a). The Bloch states at the
U-point are labeled by $\hat{G}_n$ and $\hat{S}_a$ eigenvalues ($n_{\pm}$, $a_\pm$). 
All bands are doubly degenerate, and Kramers partners $\Theta | \phi_i \rangle$ have the same eigenvalues as $|\phi_i \rangle$ on these BZ boundaries.
On the U-A and U-B lines, the   Bloch states are labeled by $n_{\pm}$. $|\phi_2 \rangle$ and $|\phi_3 \rangle$ have to be exchanged along these paths enforcing
the nodal ring FS.
}
\label{fig:tnl}
\end{figure}


\section{Topological semimetals without TRS}
In the above discussion, the TRS is crucial to ensure the four-fold degeneracy of nodal FS.  Without TRS, the twofold degeneracy for each band is no longer guaranteed. However, possible four-fold degenerate nodal FSs can still be realized in the absence of TRS, as an antiunitary operator which leads to a double degeneracy like Kramers theorem can be defined.
Below we will show how a nodal FS change under magnetic field perturbations along three different directions and identify their topological nature without TRS. 
The two degenerate states could have the same eigenvalues of non-symmorphic operations, leaving four-fold degenerate Dirac points on BZ boundary line.

Applying magnetic field along $\hat{a}$-axis breaks the $\hat{n}$-glide and mirror symmetry plane in addition to TRS. The schematic band structure are shown in Fig.~3.
Under the magnetic field along the ${\hat a}$-axis each band remains doubly degenerate on $k_b=\pi$ plane with four-fold degenerate Dirac point crossing on the U-R high symmetry line. 
An antiunitary operator $\Theta_{n}$ defined as the product of $\hat{G}_n$ and $\Theta$ leads to such degeneracy as shown below. 
\beq
\Theta_{n}\equiv\hat{G}_{n}\Theta: (x,y,z,t)\rightarrow\left(\frac{1}{2}-x,\frac{1}{2}+y,\frac{1}{2}-z,-t\right)\times I\;.
\label{eq:6a}
\eeq
While TRS and $\hat{G}_n$ are broken, $\Theta_n$ is preserved. 
Furthermore, $\Theta_{n}$ symmetry is invariant on $k_b=\pi$ plane with $\left(\Theta_{n}\right)^2=e^{i k_b} = -1$ on this BZ boundary plane. 
Therefore, two orthogonal Bloch states $|\phi\rangle$ and $\Theta_{n}|\phi\rangle$ are degenerate, similar to Kramers doublets under TRS. 
In addition, the screw axis $\hat{S}_a$ is also present.

Suppose that there is a Bloch state $|\phi\rangle$ on the U-R line with $a_+$, the $\hat{S}_{a}$ eigenvalue. Its Kramers partner $\Theta_{n}|\phi\rangle$ under $\hat{S}_{a}$ operation
shows that 
$
\hat{S}_{a}\Theta_{n}|\phi\rangle= a_+\Theta_{n}|\phi\rangle\;.
$
It carries the same screw $\hat{S}_{a}$ eigenvalue with $|\phi\rangle$.  Therefore, 
magnetic field along $\hat{a}$-direction will not lift the degeneracy along the U-R line. The four-fold degeneracy at U-point also remains intact due to the persistence of screw axis along $\hat{a}$-direction. 
This pair of Dirac nodes, as demonstrated above, is thus protected by a screw axis $\hat{S}_a$ and $\Theta_n$.
They can only be destroyed by annihilating them at BZ boundary, similar to the interlayer sublattice potential discussed in Ref. \cite{Carter12}.

\begin{figure}
\centering
\includegraphics[width=9cm]{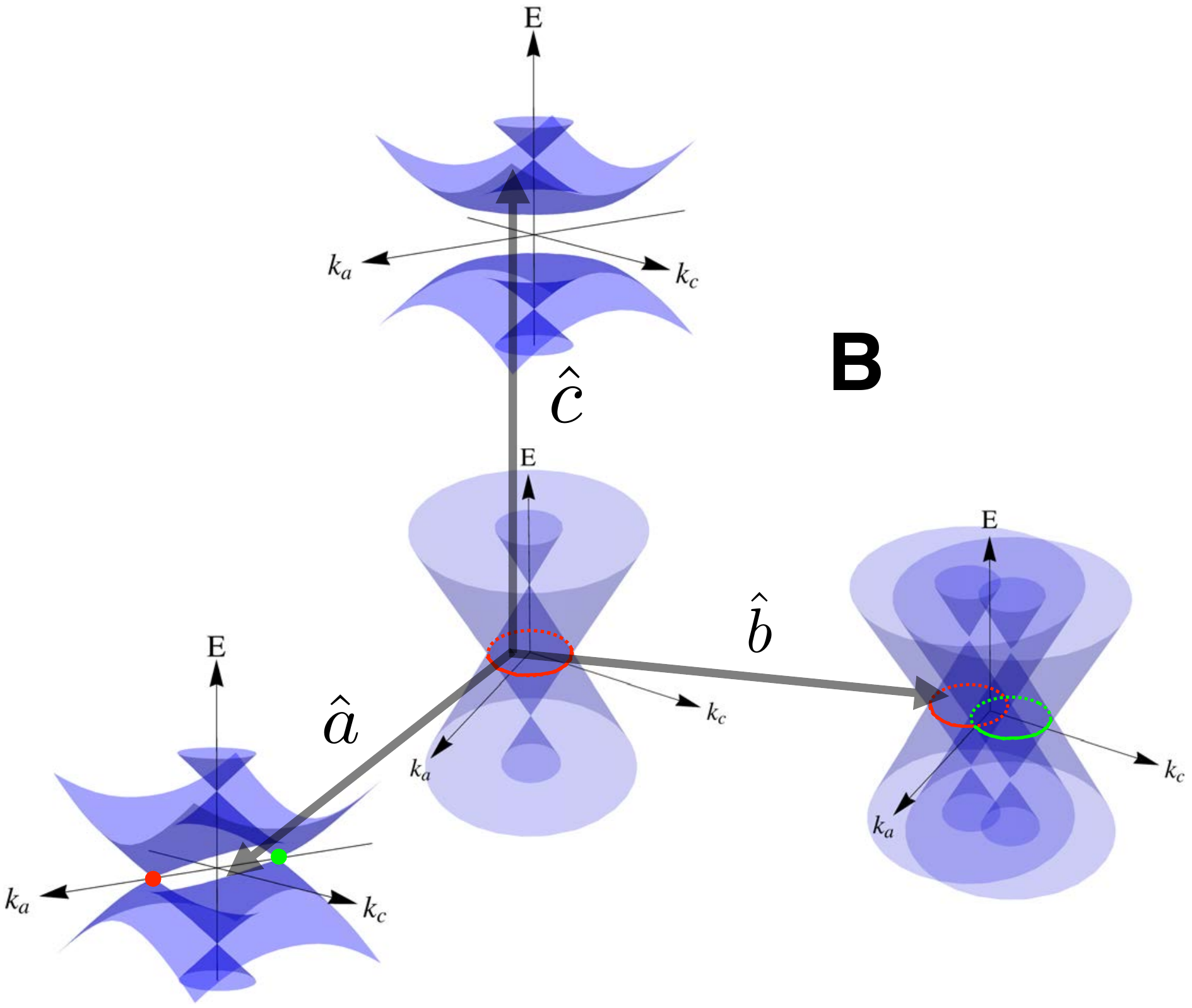}
\caption{(color online) Schematic plot to summarize how the direction of magnetic field $B$ perturbation affects the four-fold degenerate nodal-line FS. The grey axis labeled the orientation of magnetic field. All band dispersions are plotted on $k_b=\pi$ plane. The nodal line (red solid/dashed line) FS, located at the origin, represents the case when TR symmetry is preserved. When $B \parallel \hat{c}$, the nodal-line FS is destroyed by having a bulk energy gap. Two four-fold nodal FS emerge along $k_a$ direction after applying the $B$ field parallels $\hat{a}$-axis. On the other hand, the $\hat{b}$ magnetic field splits the single nodal-ring FS into two Weyl rings.}
\label{fig:0}
\end{figure}

When the field is along the ${\hat b}$-axis, $\hat{G}_b$ and TRS are both broken, and 
the four-fold degeneracy at U-point is lifted. However, the product of $\hat{G}_b$ and $\hat{T}$ are preserved on both the U-R and X-S BZ boundary lines:
\beq
\Theta_b \equiv \hat{G}_b\hat{T}:(x,y,z,t)\rightarrow \left(\frac{1}{2}-x,\frac{1}{2}+y,z,-t\right)\times i \hat{\sigma}_z.
\label{eq:b1}
\eeq
The square of this antiunitary operator $\Theta_b$ is $-1$ on $k_b=\pi$ {\it i.e.} $\Theta_b^2=e^{-i k_b}=-1$ implying that double degeneracy protected under $\Theta_b$ operation
occurs on the U-R and X-S lines.  
$\Theta_b$ operation to the Bloch state $\vert \phi_i\rangle$  with $n_+$ eigenvalues on the U-R line yields $\Theta_b|\phi_i\rangle$, which in fact possess the same n-glide eigenvalue with $|\phi_i \rangle$. 
Thus the four-fold degenerate eigenstates on the U-R line are protected by $\hat{n}$-glide and $\Theta_b$ symmetry. 
The Bloch states at R-point still remain four-fold degenerate, which can attributes to the preservation of n-glide. Meanwhile, since the screw rotation symmetry is broken by the magnetic field, a gap proportional to the strength of the magnetic field, appears at U-point. 
This degenerate Dirac nodes on the U-R line can be also understood as the intersect points between two nodal ring FSs as shown in Fig. 2. 
One four-fold degenerate nodal ring FS splits into two doubly degenerate nodal ring (Weyl ring) FSs as they move upwards and downwards, respectively, along U-X line under the presence of a magnetic field.
The overlap between two nodal ring FSs make the four-fold Dirac points, which eventually vanish when the magnetic field strength further increases.

Finally, we discuss the case with the magnetic field along the $\hat{c}$-direction, where $\hat{b}$-glide, $\hat{n}$-glide and TRS are all broken. The doubly degenerate states for each band on $k_b=\pi$ plane can be explained by another emergent antiunitary operator $\Theta_{n}$, which is defined as the product of $\hat{n}$-glide and $\Theta$ operator as in Eq.~(\ref{eq:6a}). Along U-X line, the screw rotation symmetry along $\hat{c}$-axis $\hat{S}_c \equiv \hat{G}_b\hat{G}_n$ is invariant. The degenerate pair of $\Theta_n|\phi_i\rangle$ and $|\phi_i\rangle$ on U-X line carry opposite screw eigenvalues.
Therefore, the four-fold degenerate points are avoided due to the hybridization, and completely gaped out the band degeneracy near Fermi energy. 
Similar situation also occur on the U-R line where mirror symmetry is preserved and $[\hat{M}_c,\Theta_n]=0$. It hence leads to gapped states on the U-R line. Besides, since the $\hat{n}$-glide symmetry is also broken, a generic momentum point on $k_b=\pi$ plane should have a gap, and the system turns into a trivial band insulator.

\begin{figure}
\centering
\includegraphics[width=8cm]{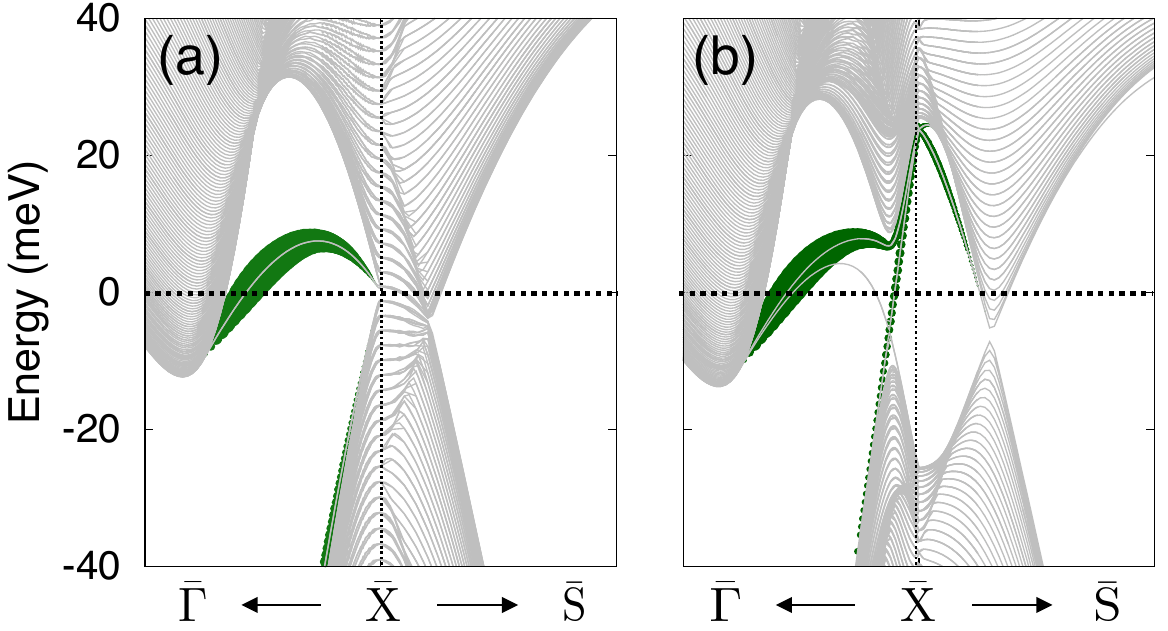}
\caption{(color online) (001) surface states in the presence of TRS.
(a) shows the bands of the 40-layer-thick (001) slab geometry, with the surface weight
of each Bloch state depicted as the size of the green symbol.
(b) shows the bands with a sub-lattice potential that breaks mirror and 
$\hat{n}$-glide symmetries. The double helical surface states are represented by the green symbols where the size refers the weight.}
\label{fig:srfc1}
\end{figure}

\begin{figure}
\centering
\includegraphics[width=8cm]{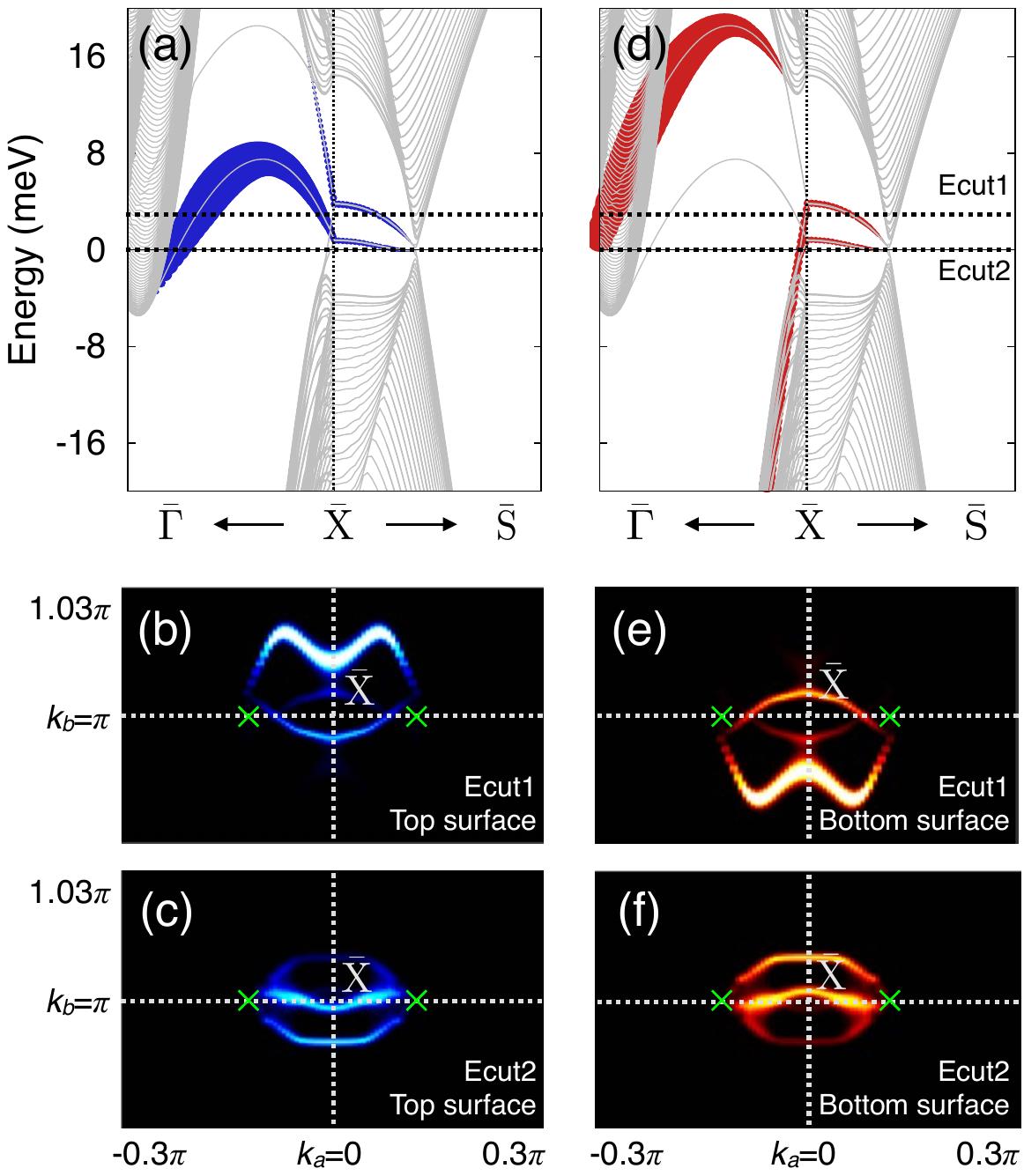}
\caption{(color online) (001) surface states 
in the presence of magnetic fields along the $\hat{a}$-direction.
(a-c) and (d-e) show the surface states on the top and bottom surfaces of the
(001) slab geometry, respectively. (b,e) and (c,e) show the constant energy cut of the
surface states in the momentum space at the energy +2.5 and 0 meV with respect to the Fermi level,
respectively, as shown in (a) and (d). Green crosses in the constant energy plots mark the
position of surface-projected bulk Dirac points.
These surface states represent two Fermi arcs from the four-fold degenerate Dirac points.}
\label{fig:srfc2}
\end{figure}

\section{Surface states}
Since the bulk nodal FSs are protected by the space-time inversion and non-symmorphic symmetries, one can ask if there are nontrivial surface states associated with the bulk states.
While the surface naturally breaks the inversion symmetry, surface states can possess non-trivial topology depending on the direction of surfaces.
Given that the double degeneracy along the U-R and X-S lines are protected by the product of the b-glide and TRS ($\Theta_b$) without involving the inversion symmetry, 
a surface containing this glide plane could be potentially interesting. The (001) surface breaks the mirror and n-glide, but preserves the b-glide symmetry.
Thus we study the (001) surface states using a tight binding model derived for perovskite iridates $A$IrO$_3$ in Ref. \onlinecite{Carter12}.

As shown in Fig. 3 (a), the (001) surface states perpendicular to $\hat{c}$-direction shows the surface bands across the $\bar{\Gamma}$-$\bar{\rm X}$ line. This is related to Z$_2$ Dirac cone discussed in Ref. \onlinecite{Heungsik15}. On the other hand, the surface states associated with the FS ring
cannot be separated from the bulk spectrum across $\bar{\rm X}$-$\bar{\rm S}$ line. 
To gap out the bulk states but keeping the $\Theta_b$ invariance, one can introduce a sub-lattice potential.\cite{Carter12,Chen15}
The surface states are then double helical states named Riemann surface states\cite{Fang15} as shown in Fig. 3 (b). 
TRS is essential for the existence of these surface states, and 
one then can ask what happens to them when the TRS is broken.
When the field is along $a$-axis, the $\hat{b}$-glide plane is still preserved and furthermore the bulk states are gapped except two nodal points.
As shown in Fig. 4, two Fermi arcs emerge from the bulk nodal points on each surface side:
the double helical state splits into two Fermi arcs, and one appears on the top surface while the other appears
on the bottom surface just like the Fermi arcs appearing from Dirac to Weyl FS. 
Unlike the conventional Dirac point discovered so far, this nodal point accompanies two Fermi arcs on each side. 
The Chern number associated with the each nodal point is $\pm 2$.

\section{Summary}
In summary, we prove that four-fold degenerate nodal line of FSs on the BZ boundary plane in 3D non-symmorphic lattices is guaranteed, 
when there are two perpendicular non-symmorphic symmetry operators, e.g. two perpendicular glide planes in addition to the space-time inversion symmetry. 
Our result is applicable for non-symmorphic crystals with perpendicular glide/screw symmetry planes. Note that, in the experimentally relevant real materials such as SrIrO$_3$, the presence of the hopping terms between the same sublattice explicitly break the chiral symmetry~\cite{Heungsik15}, and the nodal line hence acquires dispersion. While the amplitude of such hopping terms is tiny in SrIrO$_3$~\cite{Heungsik15,Chen15}, in other materials it can be a different case. However, this does not alter the main conclusion.
We also show that four-fold Dirac FSs can survive even when TRS is absent. This is because a combination of non-symmorphic and time-reversal symmetry is an antiunitary operator
which leads to the  double degeneracy like Kramers degeneracy.
Using a tight binding model derived for perovskite iridates, we also present the associated surface states with and without TRS. 
On the (001) surface where the product of the  b-glide and TRS  is preserved,
the double helical surface states are found, but they are hidden under the bulk states. When the magnetic field is applied along a certain direction that keeps the 
b-glide symmetry, the double Fermi arcs associated with four-fold Dirac point appears, which indicates that these Dirac points are made of two Weyl points with the same
topological charge.  On the other hand, a pair of Weyl ring FSs emerges under the magnetic field along another direction. 
The current work suggests that materials with non-symmorphic crystalline symmetries offer an excellent playground to explore rich topological phases.

\begin{acknowledgements}
This work was supported by the NSERC of Canada and the Center for Quantum Materials at the University of Toronto.
\end{acknowledgements}


\begin{appendix}
\section{Four-fold degeneracy enabled through non-symmorphic symmetries with space-time inversion symmtry}
Here we provide a proof for the degeneracy between a pair of Bloch states related by non-symmorphic symmetries when there are two perpendicular screw or glide symmetry operations.
The two screw operators considered in the maintext are defined in Eq.~(\ref{eq:1:a}) and (\ref{eq:1:b}),
and, note that, the screw axes are off-center from the inversion center, $(0, 0, 0)$. The axis for the $\hat{S}_a$ screw operation is parallel to $\hat{a}$-axis but it passes through $(0, b/4, 0)$, instead of $(0,0,0)$. The other screw rotation axis $\hat{S}_b$ passes through $(a/4, 0, c/4)$, and is parallel to $\hat{b}$-axis. Here $a,b$ and $c$ are the length of the Bravais lattice basis $\hat{a}, \hat{b}$ and $\hat{c}$, respectively. Note that, when squared, both $\hat{S}_a$ and $\hat{S}_b$ correctly reproduces the unit translations along $\hat{a}$ and $\hat{b}$ directions, respectively.
The space-time inversion symmetry defined as a product of TR and inversion operator, $\Theta=\hat{T}*\hat{P}$ is present:
$\Theta:\left(x,y,z,t\right)\rightarrow\left(-x,-y,-z,-t\right)\times i \hat{\sigma}_y$. The c-axis screw operator is then given by ${\hat S}_c = {\hat S}_a * {\hat S}_b$ and the glide plane operators are also found by
${\hat G}_n = {\hat S}_b * {\hat P}$ and ${\hat G}_b = {\hat S}_a * {\hat P}$.

Since the degeneracy occurs on the BZ boundary plane, let us focus on the Bloch states on $k_b=\pi$ planes.
In this plane, the Bloch states are invariant under n-glide operator $\hat{G}_n$, thus they can be classified by n-glide eigenvalues $n_{\pm}$. 
Given that 
$\left(\hat{G}_n\right)^2=-e^{i k_a + i k_c}$, $n_{\pm}=\pm i e^{i\frac{k_a+k_c}{2}}$.  All Bloch states on $k_b = \pi$ plane carry one of these eigenvalues.

Now let us exam a particular high symmetric U-point $(k_a=0, k_b=\pi, k_c=\pi)$ on this BZ boundary plane.
At U-point, the Bloch states are invariant under the screw ${\hat S}_a$ operation in addition to $G_n$. 
Thus the Bloch states at U-point are denoted by both n-glide and a-axis screw eigenvalues. 
Since $\left(\hat{S}_a\right)^2=-e^{i k_a}$, the eigenvalues of $\hat{S}_a$ is  $a_{\pm}=\pm i e^{i \frac{k_a}{2}}$which is $\pm i$ at U-point.
Taking the b-glide operation on a Bloch state $|\phi_1\rangle$, i.e., ${\hat G}_b |\phi_1 \rangle$ generates another Bloch state with the same n-glide eigenvalue
but different screw ${\hat S}_a$ eigenvalue.  These two are orthogonal and degenerate. The proof is shown as follows. 

Let's consider a Bloch state $|\phi_1\rangle$ which carries $n_+$ and $a_+$ eigenvalues.  
Note that 
under $\hat{G}_n$ and $\hat{S}_a$, $\hat{G}_b|\phi_1\rangle$ behave as
\bea
\hat{G}_n\left(\hat{G}_b|\phi_1\rangle\right)&=&\hat{G}_b\left(\hat{G}_n|\phi_1\rangle\right)=n_+\hat{G}_b|\phi_1\rangle\\
\hat{S}_a\left(\hat{G}_b|\phi_1\rangle\right)&=&-\hat{G}_b\left(\hat{S}_a|\phi_1\rangle\right)=-a_+\hat{G}_b|\phi_1\rangle\nonumber\\
&=&a_-\hat{G}_b|\phi_1\rangle,
\label{eq:a7}
\eea
where we used the commutation relations given in Table in the next section: $\hat{G}_b$ commutes with $\hat{G}_n$ but anticommutes with $\hat{S}_a$. We also used $a_-=-a_+$. 
This suggests that $\hat{G}_b|\phi_1\rangle$ is also an eigenstate of both ${\hat G}_n$ and $\hat{S}_a$ operators with $n_+$ and $a_-$ eigenvalues, respectively. As mentioned in the maintext, 
we denote this Bloch state $|\phi_3 \rangle$ which is proportional to  ${\hat G}_b |\phi_1 \rangle$ up to U(1) phase factor.
Furthermore, the inner product of these two Bloch states is given by
\bea
\langle \phi_1| \hat{G}_b|\phi_1\rangle&=&-\langle \phi_1| \left(\hat{S}_a\right)^2\hat{G}_b|\phi_1=-\left(-\hat{S}_a|\phi_1\rangle\right)^{\dagger}\hat{S}_a\hat{G}_b|\phi_1\rangle\nonumber\\
&=&-\langle \phi_1|\hat{G}_b|\phi_1\rangle,
\label{eq:a8}
\eea
where $\left(\hat{S}_a\right)^2=-1$ at U-point is used. This implies $\langle \phi_1|\hat{G}_b|\phi_1\rangle=0$ and thus $|\phi_1\rangle$ and $|\phi_3\rangle$ are orthogonal. 
Since $\hat{G}_b$ commutes with the Hamiltonian, we prove that $|\phi_1\rangle$ and $|\phi_3\rangle$ are degenerate at U-point. 
Following the similar argument, another pair of Bloch states $(|\phi_2\rangle,|\phi_4\rangle)$ are degenerate where $|\phi_4\rangle \propto \hat{G}_b|\phi_2\rangle$.
Thus taking into account their Kramers partners\cite{FangPRB15}, two sets of four Bloch states -- 
$\left(|\phi_1\rangle,|\phi_3\rangle,\Theta| \phi_1\rangle,\Theta|\phi_3\rangle\right)$ and $\left(|\phi_2\rangle,|\phi_4\rangle,\Theta|\phi_2\rangle,\Theta|\phi_4\rangle\right)$ -- are degenerate at U-point.

How do these Bloch states evolve as they move to a generic point on R-S and X-S BZ boundary line in the $k_b =\pi$ BZ plane?
As discussed in the maintext, along R-S and X-S BZ boundary line,   $|\phi_1\rangle$ and $|\phi_2\rangle$ ($|\phi_3\rangle$ and $|\phi_4\rangle$) are degenerate and related by $\hat{G}_b$ ($\hat{S}_a$).
To prove our statement, let us consider an arbitrary point on R-S line $(k_a=\pi,k_b=\pi)$, where the Bloch states are invariant under the b- and n-glide operation.  Using the commutation relation given in Table below, i.e., 
$ \hat{G}_b\hat{G}_n=-e^{-i(k_a=\pi)+i(k_b=\pi)}\hat{G}_n\hat{G}_b=-\hat{G}_n\hat{G}_b$,
the Bloch state $\hat{G}_b |\phi_1\rangle$ on the R-S line under n-glide operator $\hat{G}_n$ carries $n_-$ eigenvalue as shown below. 
\beq
\hat{G}_n\left(\hat{G}_b|\phi_1\rangle\right)=-\hat{G}_b\left(\hat{G}_n|\phi_1\rangle\right)=-n_+\hat{G}_b|\phi_1\rangle=n_-\hat{G}_b|\phi_1\rangle.
\label{eq:a10}
\eeq
Thus $\hat{G}_b|\phi_1\rangle$ with opposite n-glide eigenvalue becomes degenerate with $|\phi_1\rangle$ along the R-S line, as $\hat{G}_b$ commutes with the Hamiltonian
on the R-S line. 
Using the similar process  including orthogonality, $\langle \phi_1|\hat{G}_b|\phi_1\rangle=\langle \phi_1|\hat{G}_b\Theta|\phi_1\rangle=0$,
we showed that two pairs of four Bloch states, $\left(|\phi_1\rangle, |\phi_2\rangle, \Theta |\phi_1 \rangle, \Theta |\phi_2 \rangle \right)$ and $\left(|\phi_3\rangle,|\phi_4\rangle, \Theta |\phi_3 \rangle,
\Theta |\phi_4 \rangle \right)$ including Kramers doublet are degenerate on R-S line.
Hence $|\phi_2\rangle$  and $|\phi_3 \rangle$ with opposite  and same n-glide eigenvalue at U-point exchange their degenerate parter, as they move from U to any other point along the R-S line
as shown in Fig. 1 (b) in the main text. The energy level crossing should occur somewhere in between, unless these non-symmorphic symmetries are broken. 

Similarly, a band crossing should occur somewhere between U-point to any point on the X-S  BZ boundary line $(k_b=\pi,k_c=0)$,
where the Bloch states are invariant under ${\hat G}_n$ and $\hat{S}_a$ operations. Since they anticommute, i.e.,
$ \hat{S}_a\hat{G}_n=-e^{-i (k_c=0)} \hat{G}_n \hat{S}_a = -\hat{G}_n\hat{S}_a$, another Bloch state $\hat{S}_a|\phi_1\rangle$ generated by
taking $\hat{S}_a$ on $|\phi_1\rangle$ has the opposite n-glide eigenvalue from $|\phi_1\rangle$.
 Since  $|\phi_2 \rangle \propto {\hat S}_a |\phi_1\rangle$ on the X-S line,  and ${\hat S}_a$ commutes
with the Hamiltonian,  we find $\left(|\phi_1\rangle, |\phi_2\rangle, \Theta |\phi_1 \rangle, \Theta |\phi_2 \rangle \right)$ and $\left(|\phi_3\rangle,|\phi_4\rangle, \Theta |\phi_3 \rangle, \Theta |\phi_4 \rangle \right)$ are degenerate on the X-S line. 

In summary, we prove that two pairs of Bloch states denoted as $(|\phi_1\rangle,|\phi_3\rangle)$ and $(|\phi_2\rangle,|\phi_4\rangle)$ at U-point must switch a degenerate partner when the bands move from U point towards the BZ boundary line of X-S and R-S line, which results in a ring of four-fold degenerate FS on $k_b =\pi$ BZ plane.

The commutation table among b-glide, n-glide, mirror and $\Theta$ operators established in the end is used to demonstrate the mathematical proof provided in the above discussion. 

\begin{widetext}
\beq
\begin{tabular}{|c||c|c|c|c|}
\hline
Symmetry & $\hat{G}_b$ & $\hat{G}_n$ & $\hat{M}_c$ & $\Theta$ \\ \hline
$\hat{G}_b$ & 0 & $\hat{G}_b\hat{G}_n=-e^{-i k_a+i k_b} \hat{G}_n\hat{G}_b$ & $\{\hat{G}_b,\hat{M}_c\}=0$ & $\hat{G}_b\Theta=e^{-i k_a+ik_b}\Theta \hat{G}_b$  \\ \hline
$\hat{G}_n$ & $\hat{G}_n\hat{G}_b=-e^{i k_a -i k_b} \hat{G}_b\hat{G}_n$ & 0 & $\hat{G}_n\hat{M}_c=-e^{i k_c} \hat{M}_c\hat{G}_n$ & $\hat{G}_n\Theta=e^{i k_a - ik_b + ik_c}\Theta \hat{G}_n$ \\ \hline
$\hat{M}_c$ & $\{\hat{G}_b,\hat{M}_c\}=0$ & $\hat{M}_c\hat{G}_n=-e^{-i k_c} \hat{G}_n\hat{M}_c$  & 0 & $\hat{M}_c\Theta=e^{-i k_c}\Theta \hat{M}_c$ \\ \hline
$\Theta$ &$\hat{G}_b\Theta=e^{-i k_a+ik_b}\Theta \hat{G}_b$  & $\hat{G}_n\Theta=e^{i k_a - ik_b + ik_c}\Theta \hat{G}_n$ & $\hat{M}_c\Theta=e^{-i k_c}\Theta \hat{M}_c$ & 0 \\ \hline
\end{tabular}
\label{eq:tab1}
\eeq
\end{widetext}

\end{appendix}


\begin{thebibliography}{99}
\bibitem{foot1} We use the energy cuts in the band structures shown in Ref.~\cite{Heungsik15}.
\end{thebibliography}
\end{document}